\documentclass[pdftex]{article}
\usepackage{icrctc07}

\title{Observation of the binary system LS I +61 303 in Very-High Energy Gamma-Rays with VERITAS}
\shorttitle{Observation of LS I +61 303 with VERITAS}
\authors{Gernot Maier$^{1}$ for the VERITAS Collaboration$^{2}$ }
\shortauthors{G.Maier et al.}
\afiliations{$^{1}$ McGill University, 3600 University Street, H3A 2T8 Montreal, QC, Canada, \\
             $^{2}$ For full author list see G.Maier, ``VERITAS: Status and Latest Results", these proceedings}
\email{maierg@physics.mcgill.ca}

\abstract{
The high mass X-ray binary LS I +61 303 has been observed over several months in 2006 and 2007
with the \mbox{VERITAS} array of imaging air-Cherenkov telescopes. 
A signal of high energy gamma rays with energies above 350 GeV
is detected in several orbital cycles of the binary system.
The detected flux of gamma rays is strongly variable with the orbital period of 26.5 days,
while the maximum flux (corresponding to about 10\% of the flux of the Crab Nebula)
is always found at approximately apastron, suggesting a strong dependence of particle acceleration and/or propagation on the
relative position of the two objects in the system.
}

\begin{document}
\maketitle

\vspace{-0.35cm}
\section{Introduction}
\vspace{-0.30cm}

The high-mass X-ray binary system \mbox{LS I +61 303}
is one of only three X-ray binaries detected in high-energy gamma rays \cite{Aharonian05,Aharonian06,Albert06}.
It consists of a massive
Be-type star surrounded by a dense circumstellar disk
and a compact object (neutron star or black hole).
Optical observations show that the compact object orbits the star every 26.5 
days \cite{Gregory02} on a close orbit,
characterized by a semi-major axis of a few stellar radii only.
The periastron takes place at phase 0.23, apastron is at phase
0.73, inferior and superior conjuctions are at phases 0.26 and 0.16
(phase zero defined from radio observations \cite{Casares05}).
\mbox{LS I +61 303} shows highly variable emission depending on the orbital
phase across all wavelengths from radio \cite{Gregory02} to TeV gamma-rays \cite{Albert06}.

The unknown nature of the compact object and the uncertainty in the 
geometry of the system allows the existence of at least two possible models
for the origin of the high-energy emission from \mbox{LS I +61 303}.
The first class of models explains the emission of $\gamma$-rays
through the production of non-thermal particles in an
accretion powered relativistic jet (``microquasar model") \cite{Taylor84,Mirabel94}.
In the second class of models, particles are accelerated in the shock created
by the collision of a relativistic
pulsar wind with the wind of the companion star \cite{Maraschi81}. 
In both cases, the $\gamma$-ray emission is interpreted as inverse Compton (IC) scattering
of stellar radiation by high-energy electrons.
Alternative models of $\gamma$-ray production include IC e$^{\pm}$ pair cascades in the field
of massive stars \cite{Bednarek06} or $\pi^{0}$ production and decay in
hadronic interactions \cite{Romero05}.

This paper reports on the results from observations of \mbox{LS I +61 303} with VERITAS; 
a study of the correlation between these data and contemporaneous X-ray measurements can be found in a complementary
contribution \cite{AS07}.

\vspace{-0.35cm}
\section{Observations with VERITAS}
\vspace{-0.30cm}

VERITAS is an array of four imaging Cherenkov telescopes located at the Fred Lawrence
Whipple Observatory in southern Arizona \cite{RO07}.
The construction of VERITAS is now complete, but was in progess 
during the observations of \mbox{LS I +61 303} described here.
Observations were made with two telescopes during the period from September to November 2006 (Telescopes 1 and 2),
and with three telescopes during January and February 2007 (Telescopes 1, 2, and 3).
The sensitivity for $\gamma$-rays with energies above 230 GeV
of the two-telescope array in the configuration of Autumn 2006
(5 $\sigma$ in 3.3 h for a source with 10\% of the flux of the Crab Nebula
observed at $70^{\mathrm{o}}$ elevation)
was somewhat lower than that of the three-telescope array operating in early 2007
(5 $\sigma$ in 1.2 h for a similar source).
A detailed description of the telescopes, data acquisition, and calibration techniques of VERITAS
can be found elsewhere \cite{RO07,JH05,DH07}.

The LS I +61 303 observations consist of 44 hours of data (32 h of 2 telescope and 12 hours of 3 telescope data, after run quality selection),
taken at a mean elevation of  around $60^{\mathrm{o}}$.
All observations were taken in wobble mode with offsets of the pointing direction from the position
of \mbox{LS I +61 303} in the sky of $0.3^{\mathrm{o}}$ for the
2006 data and  $0.5^{\mathrm{o}}$ for the 2007 data.
Observations were made only when the moon was below the horizon. As a result of this, and the fact that 
the length of an orbital cycle of \mbox{LS I +61 303} is very close to the moon cycle, no observations are available in this data set at orbital phases between 0.95 and 0.2.

The data have been analysed using independent analysis packages, several cosmic-ray rejection methods, and different
procedures to calculate the background rate at the position of the potential source (see \cite{MD07} for details
on the analyses).
All of these analyses yield consistent results.

\vspace{-0.35cm}
\section{Detection of LS I +61 303}
\vspace{-0.30cm}
%
%
%
\begin{table*}
\begin{center}
\begin{tabular}{l|c|c|c|c|c|c|c|c}
Month & Config- & Wobble & Obs.time & On     & Bck.   & Norm. & Signi-  & Flux \\
      & uration & offset&  [h]     & events & events &       & ficance & [Crab units] \\
\hline
September 2006 & 2-tel  & 0.3$^{\mathrm{o}}$ & 7.2 & 396 & 859 & 0.33 & 5.2$\sigma$ & 0.064$\pm$0.01 \\
October 2006   & 2-tel  & 0.3$^{\mathrm{o}}$ & 7.0 & 420 & 794 & 0.33 & 7.4$\sigma$ & 0.092$\pm$0.01 \\
November 2006  & 2-tel  & 0.3$^{\mathrm{o}}$ & 5.0 & 228 & 498 & 0.33 & 3.9$\sigma$ & 0.050$\pm$0.01 \\
January 2007   & 3-tel  & 0.5$^{\mathrm{o}}$ & 3.6 & 178 & 507 & 0.30 & 3.8$\sigma$ & 0.049$\pm$0.01  \\
February 2007  & 3-tel  & 0.5$^{\mathrm{o}}$ & 2.3 & 120 & 394 & 0.30 & 1.9$\sigma$ & $<$0.065 \\
\end{tabular}
\caption{\label{tab:observations} Summary of observations of LS I +61 303 during the high state orbital phases from 0.5 to 0.8.
Upper limits on the integral $\gamma$-ray flux are given for significances below 3 $\sigma$ for events
above an energy of 350 GeV assuming a differential spectral index of the source spectrum of $\Gamma=-2.5$.
The confidence level is 99\%.
}
\end{center}
\end{table*}

\begin{figure}
\centering
\noindent
%
%
\includegraphics*[width=0.46\textwidth]{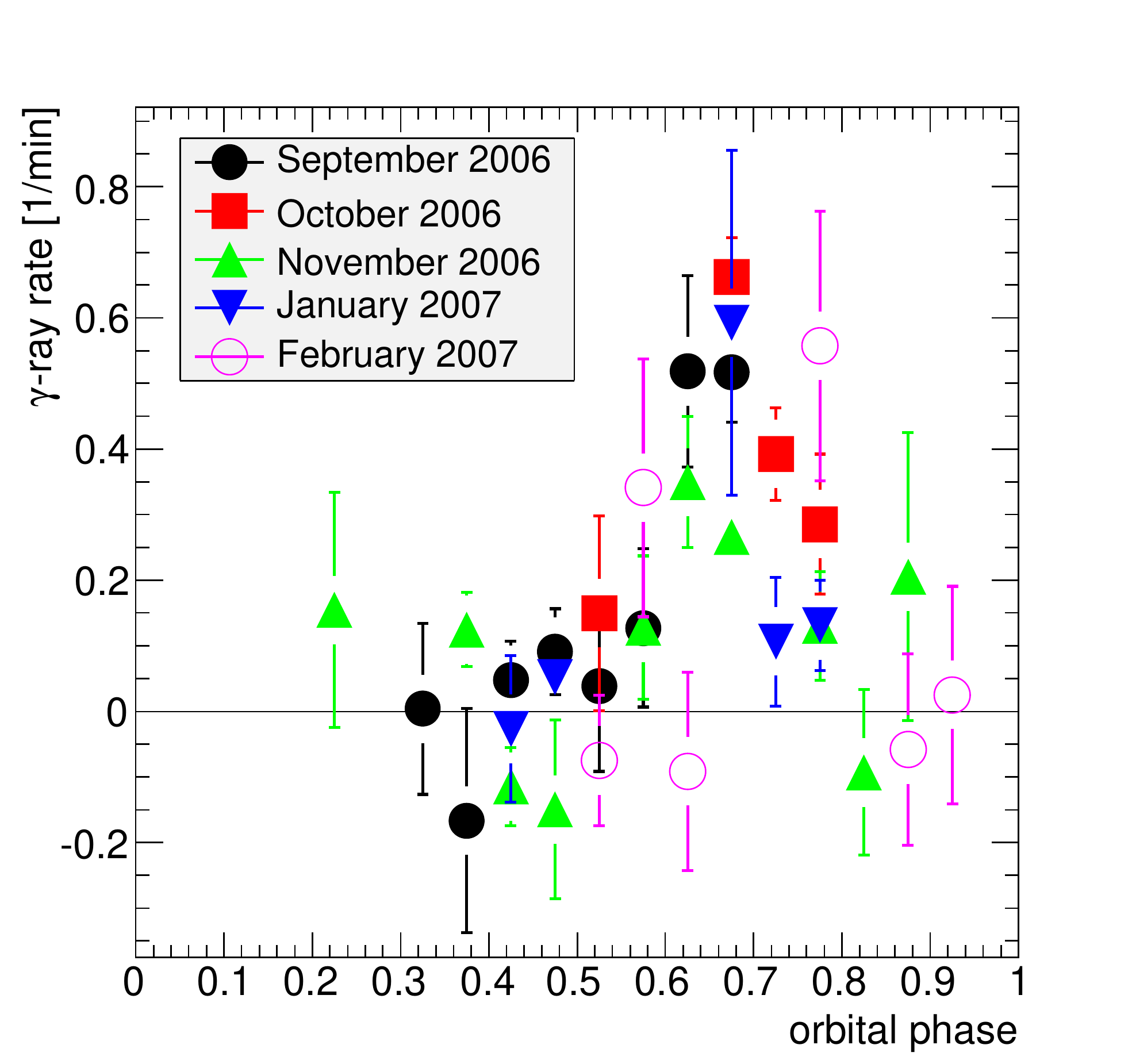}
\caption{\label{fig1}
Rate of excess events versus orbital phase for five different orbits.
Rates are not corrected for different elevations.
}
\end{figure}

\begin{figure}
\centering
%
%
%
%
\includegraphics*[width=0.46\textwidth]{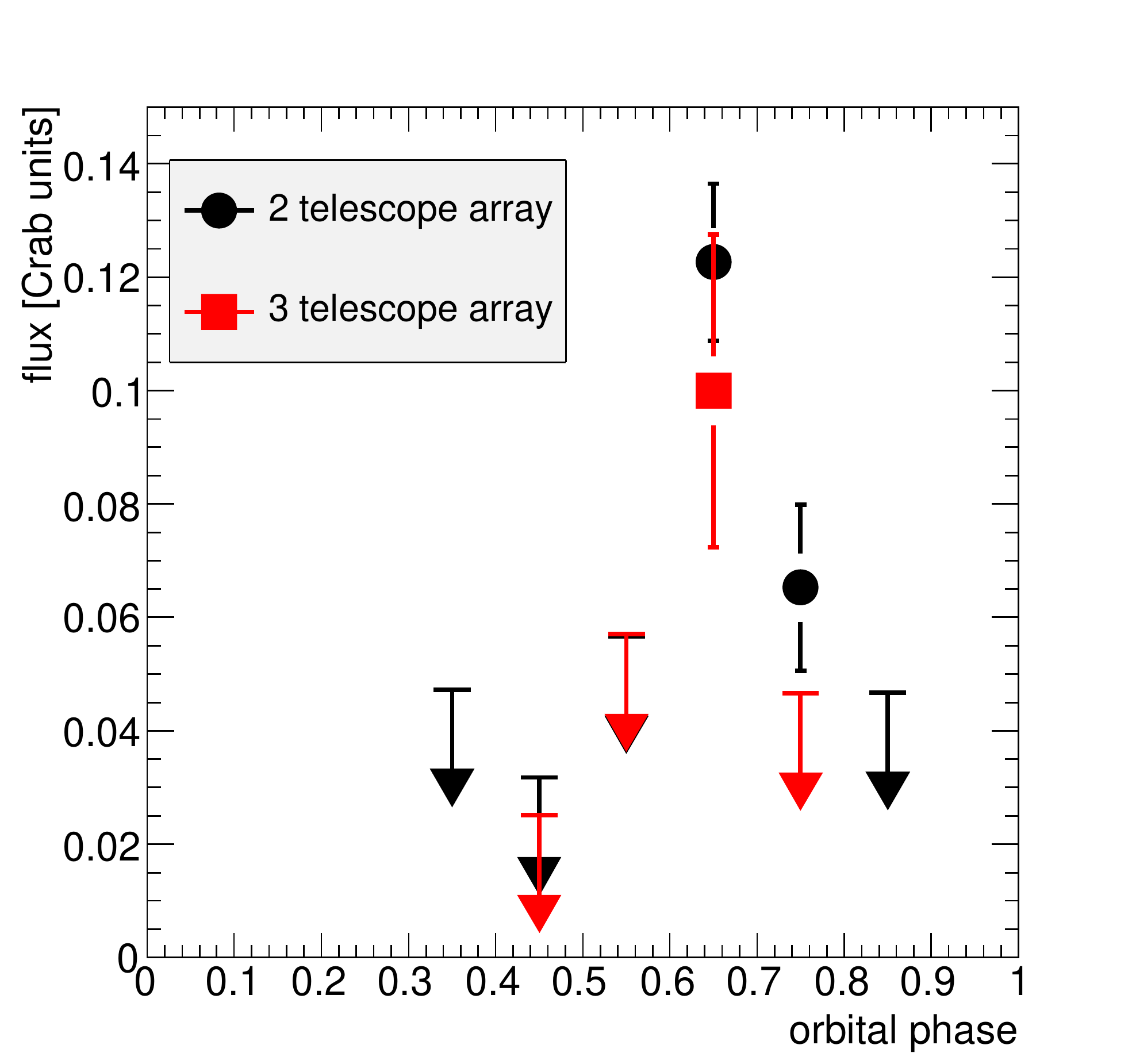}
\caption{\label{fig2} Fluxes and upper flux limits in units of the flux of the Crab Nebula versus orbital phase
from observations of \mbox{LS I +61 303}. 
A significance of $3\sigma$ is required for all flux calculations; upper flux limits are 99\% confidence limits.
}
\end{figure}

\mbox{LS I +61 303} has been detected by VERITAS 
with a total significance of 8.8 $\sigma$.
The emission of photons from \mbox{LS I +61 303}
is, as previous observations at these energies indicate \cite{Albert06}
and studies at other wavelengths show (e.g.\cite{Grundstrom07} and references therein), strongly variable with orbital phase.
Figure \ref{fig1} shows the measured  $\gamma$-ray rate vs orbital phase for different orbital cycles.
The rates are not corrected for any observational effects like elevation, array configuration, or dead times (3-7\%),
in order to illustrate how apparent the variability is.
$\gamma$-ray emission is only measurable when the two objects of the binary system are furthest away from each other 
(radio phases 0.5 to 0.8).
Table \ref{tab:observations} summarizes the observations during these high state orbital phases.
The source has been detected at a significance level of about 4 $\sigma$ or above during all except one orbital cycle (February 2007).
The lack of a detection in this cycle is consistent with measurements in other cycles due to the small exposure during this month. 
Additionally, the data in this month were not taken at orbital phases close to 0.7,
at which the highest rates are measured during other cycles.
Combining the data from all five observed orbits shows that the flux of high energy \mbox{$\gamma$-rays} 
varies between lower than 3\% of the Crab Nebula flux (using a confidence limit of 99\% and assuming
a Crab-Nebula-like spectral energy distribution) to above 10\% in the high state
phase\footnote{
The relatively low flux makes it impossible to sample the data in shorter intervals than the 
width of the high state.
The absolute level of the flux in each bin therefore depends strongly on the definition of the phase intervals.
Varying starting points and lengths of the phase intervals can alter flux values by up to 50\%.
}
(see Figure \ref{fig2}).
The combination of data from different orbital cycles assumes implicitly that
the maximum emission takes place at the same phase during every orbit and that the fluxes do not vary from
one cycle to the next.
While the available data appear to indicate that this is the case (Figure \ref{fig1}), the statistics are sufficiently poor that the existence of significantly varying lightcurves cannot be excluded
and are even predicted in some emission models (e.g.\cite{Neronov05}).
In addition, the folding of the data with the length of the radio orbit assumes similar
periodicity for the radio and the $\gamma$-ray data. 
Using Lomb-Scargle statistics \cite{Scargle82},
the probability for a period of 26.4960 days \cite{Gregory02} in the $\gamma$-ray data is 99.94\%.

\begin{figure}
\centering
%
%
%
\includegraphics*[width=0.46\textwidth]{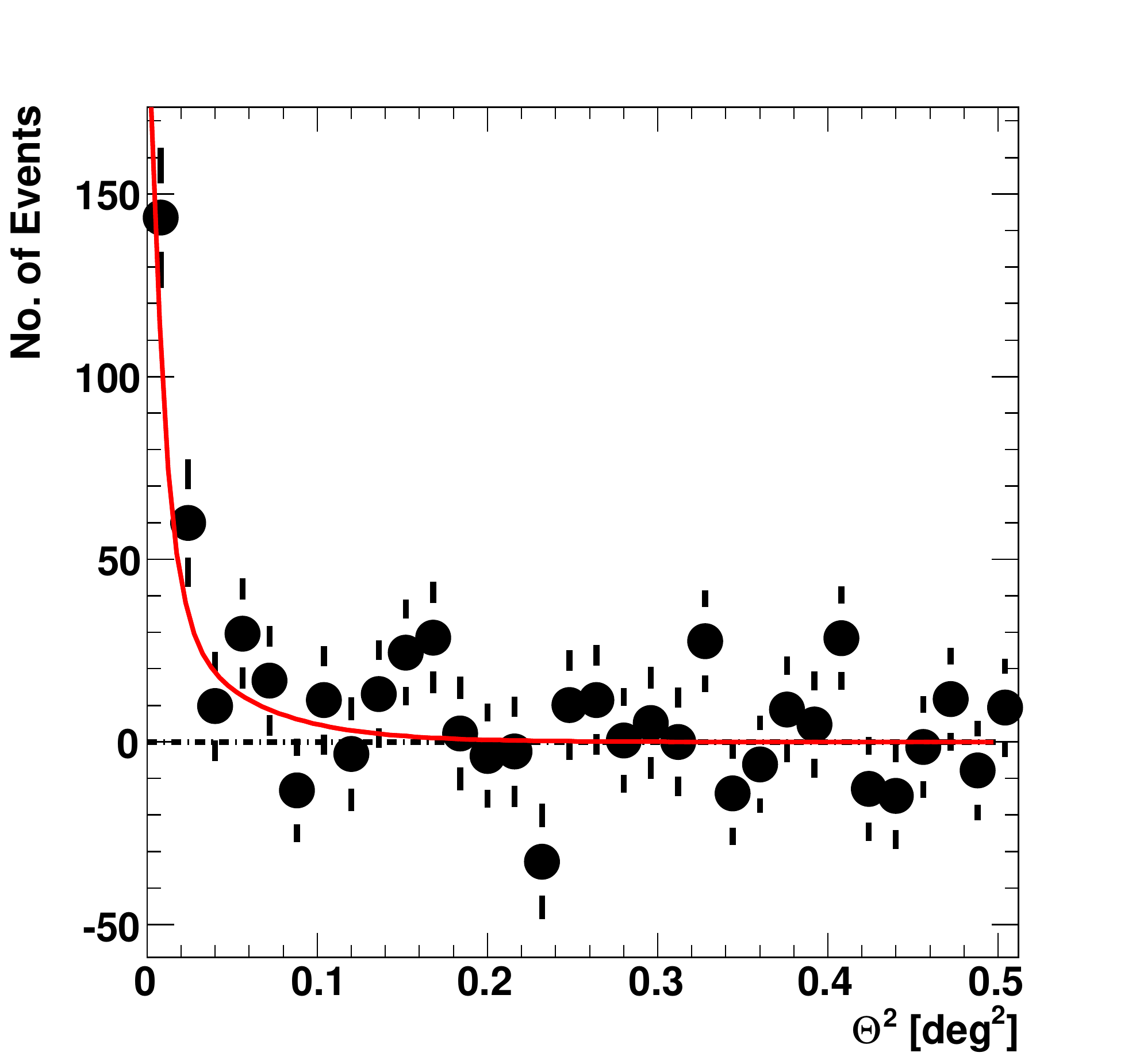}
\caption{\label{fig9}
The distribution of $\Theta^2$ for excess events (points)
from \mbox{LS I +61 303} during orbital phases 0.6-0.7 and
from Monte Carlo simulations assuming a point-like source.}
\end{figure}

\begin{figure*}
\centering
%
%
%
\includegraphics*[width=0.46\textwidth]{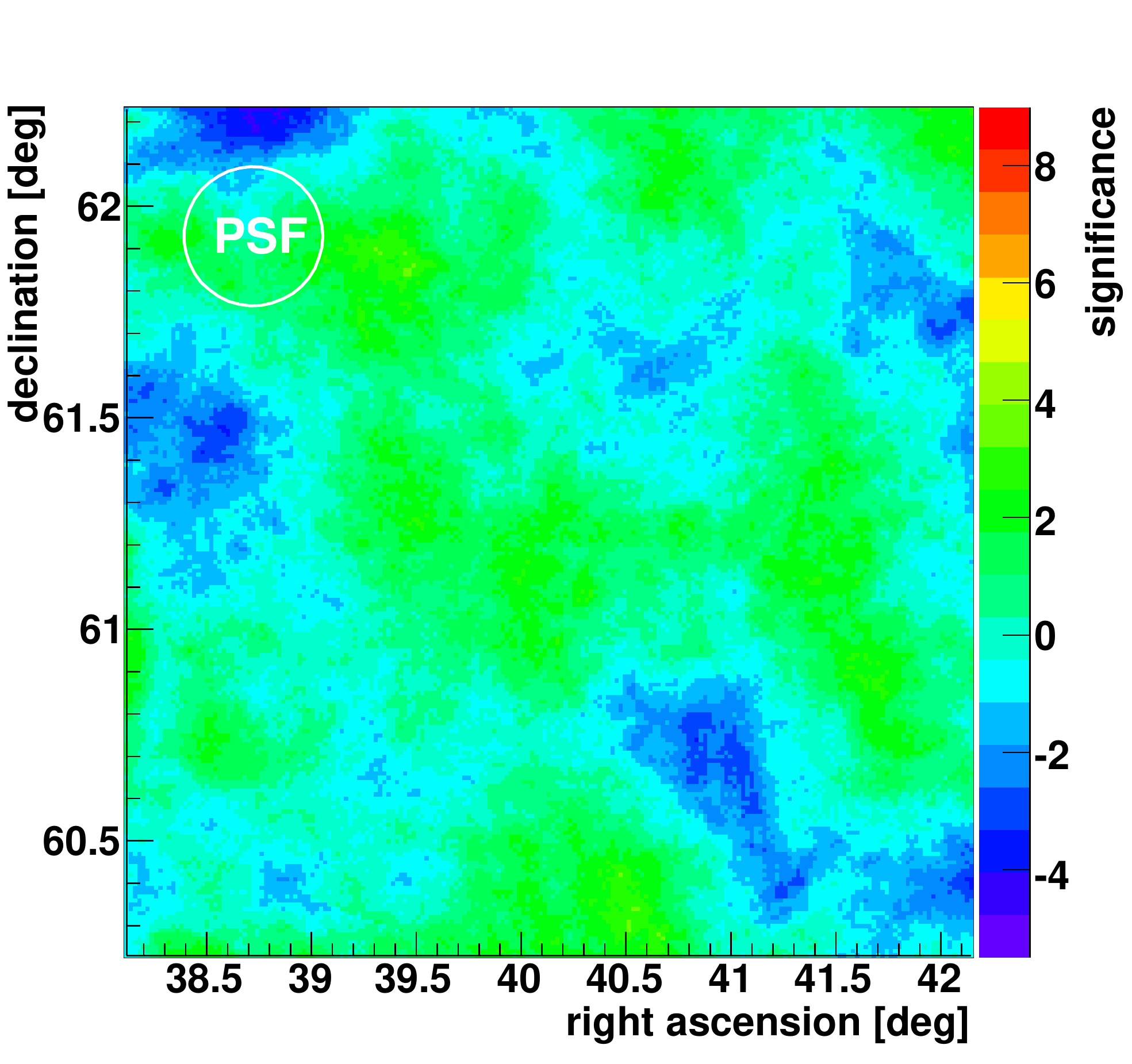}
\includegraphics*[width=0.46\textwidth]{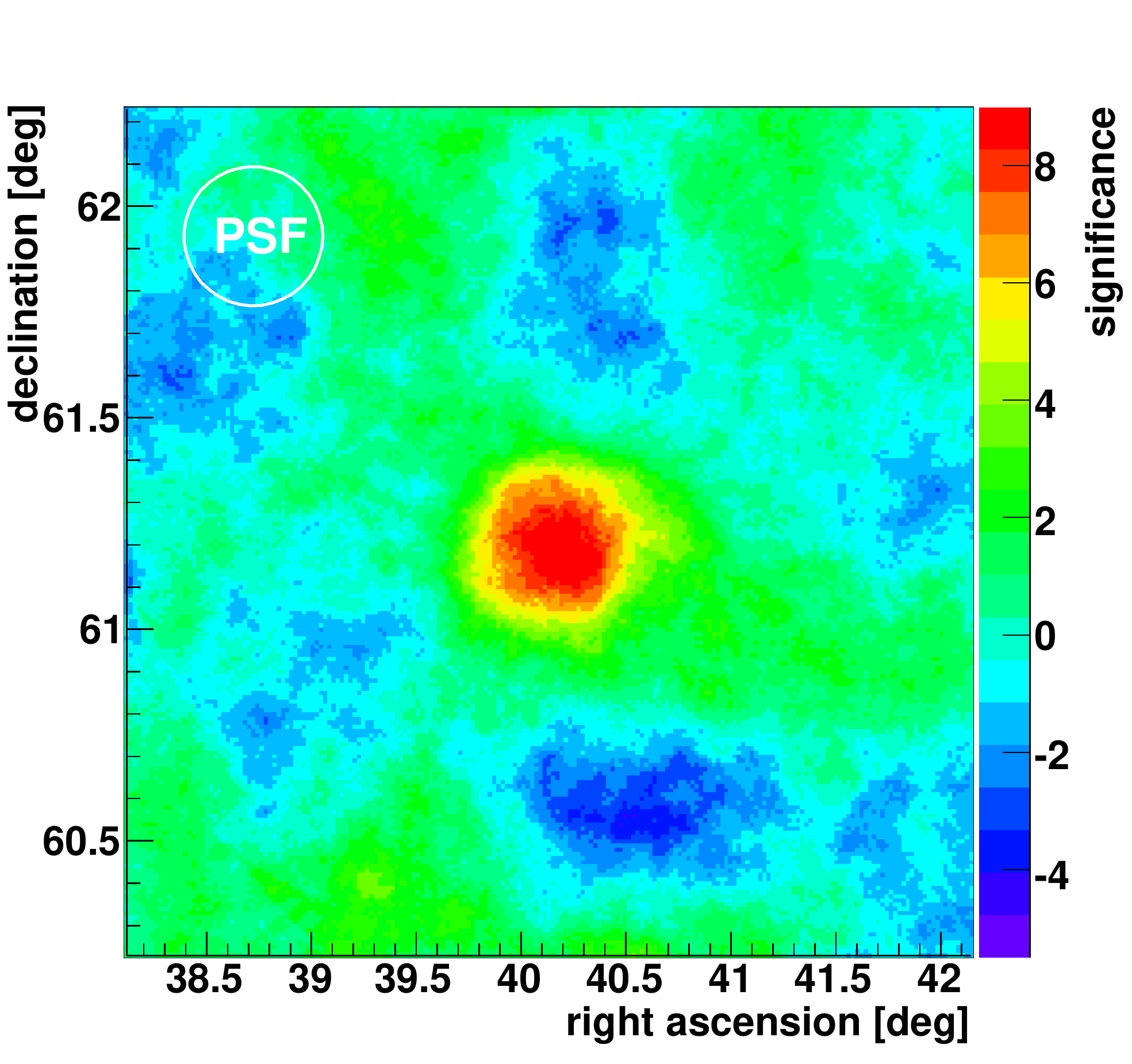}
\caption{\label{fig3} Significance map of the region around LS I +61 303 for data taken with two telescopes only.
Left: Observations during orbital phases 0.8 to 0.5 (about 13 hours of data). 
Right: Observations during orbital phases 0.5 to 0.8 (about 18 hours of data).
The white circle shows the 1 $\sigma$ point spread function for the two-telescope array of VERITAS.
}
\end{figure*}

In order to study the position and extension of the $\gamma$-ray emission around \mbox{LS I +61 303},
a sky map of the significance of the signal around the binary system was constructed (see Figure \ref{fig3}).
The extension of the excess is compatible with the point spread function of the two telescope system
and therefore consistent with a point source (Figure \ref{fig9}).
The maximum in the sky map, determined by a fit of a two dimensional Gaussian distribution to the image, 
is compatible with the position of the optical counterpart.

\vspace{-0.35cm}
\section{Conclusions}
\vspace{-0.30cm}

The detection of high-energy emission from \mbox{LS I +61 303} by VERITAS
shows strong variability which is clearly linked to the orbital motion of the binary system.
The majority of the $\gamma$-ray flux is emitted at apastron over a time scale of a only a few days.
This suggests that the $\gamma$-ray production region is close to or inside the
binary system and that the tight orbit, combined with 
the dense stellar wind of the Be-star, produces a continuously
changing environment for particle acceleration and absorption.
Future observations with the complete four-telescope VERITAS array will allow us to test the stability of the lightcurve over several orbits and to measure the emission spectrum as a function of orbital phase. These measurements, along with further contemporaneous multiwavelength observations, will provide constraining tests of the available models.

\vspace{-0.35cm}
\subsection*{Acknowledgments}
\vspace{-0.30cm}

This research is supported by grants from the U.S. Department of Energy,
the U.S. National Science Foundation,
and the Smithsonian Institution, by NSERC in Canada, by PPARC in the UK and
by Science Foundation Ireland.


\end{document}